# Surveillance of COVID-19 Pandemic using Social Media: A Reddit Study in North Carolina

Christopher Whitfield
Computer Science
North Carolina A&T University
Greensboro NC USA
clwhitfield@aggies.ncat.edu

Yang Liu
Computer Science
North Carolina A&T University
Greensboro NC USA
yliu2@aggies.ncat.edu

Mohd Anwar
Computer Science
North Carolina A&T University
Greensboro NC USA
manwar@ncat.edu

## ABSTRACT

Coronavirus disease (COVID-19) pandemic has changed various aspects of people's lives and behaviors. At this stage, there are no other ways to control the natural progression of the disease than adopting mitigation strategies such as wearing masks, watching distance, and washing hands. Moreover, at this time of social distancing, social media plays a key role in connecting people and providing a platform for expressing their feelings. In this study, we tap into social media to surveil the uptake of mitigation and detection strategies, and capture issues and concerns about the pandemic. In particular, we explore the research question, *"how much can be learned regarding the public uptake of mitigation strategies and concerns about COVID-19 pandemic by using natural language processing on Reddit posts?"* After extracting COVID-related posts from the four largest subreddit communities of North Carolina over six months, we performed NLP-based preprocessing to clean the noisy data. We employed a custom Named-entity Recognition (NER) system and a Latent Dirichlet Allocation (LDA) method for topic modeling on a Reddit corpus. We observed that *mask*, *flu*, and *testing* are the most prevalent named-entities for "Personal Protective Equipment", "symptoms", and "testing" categories, respectively. We also observed that the most discussed topics are related to *testing*, *masks*, and *employment*. The mitigation measures are the most prevalent theme of discussion across all subreddits.

## 1 Introduction

The COVID-19 disease has led to a global pandemic and the latest global public health crisis. At the time of writing this paper, global cases have surpassed 172 million (over 33 million US cases), and there are more than 3 million deaths worldwide (over 597 thousand US deaths)[1]. As a result, governments worldwide have developed new policies and mandates to help mitigate the spread of the disease. Specifically, Centers for Disease Control and Prevention (CDC) recommended wearing a mask, washing hands, and watching social distance. These recommendations are also implemented by various state and city ordinances.

Although research shows that the mitigation strategies are effective to flatten the transmission curve, public uptake of these strategies makes a huge difference. However, there is a gap in understanding the public uptake of mitigation measures or public concerns due to the pandemic. At this time of social isolation, social media has become the platform of choice for many people to express themselves which can be a valuable resource for study. Social media platforms have a strong worldwide presence of approximately 3.6 billion users as of January 2020 which accounts for nearly 49% of the global population per statistica.com.

During times of public health crises, social media platforms such as Facebook, Twitter, and Reddit are inundated with people's perceptions, opinions, and concerns. Consequently, large data sources are created that allow researchers opportunities to mine meaningful information using various methodologies. Increasingly, researchers are using Natural Language Processing (NLP) methods such as topic modeling and Named-Entity Recognition to mine public health-related data. Similarly, we leveraged the publicly available posts from Reddit, a popular social media platform, to surveil public uptake and concerns relating to the mitigation measures and burden of the disease.

In this study, we collected posts from location-specific subreddits, micro-communities within Reddit platform, as a data source to monitor the COVID-19 pandemic in North Carolina. We employed NLP techniques, particularly LDA topic modeling and custom NER, to capture issues and concerns and examine the prevalent mitigation strategies about the pandemic. More specifically, we wanted to answer the research question *"how much can be learned regarding the public uptake of mitigation strategies and concerns about COVID-19 pandemic by using natural language processing on Reddit posts?"*. The main contributions of this paper are as follows:

- We built a comprehensive cleaned corpus of COVID-19 pandemic-related posts from North Carolina subreddit communities using various NLP techniques.
- We trained and evaluated an extended, custom NER model to assess the uptake of mitigation measures against the spread of COVID-19 disease.
- We extracted people's concerns about the pandemic using an LDA-based topic model, revealing interesting side topics related to the pandemic.

---
[1] https://coronavirus.jhu.edu

The rest of this paper is organized as follows. Section 2 outlines work related to this study. In Section 3, we present methods for constructing datasets and NLP-based experiments. Section 4 presents the results and limitations of this study, and in Section 5, we conclude and outline future work.

## 2 Related Work

Considerable research has increasingly demonstrated the effectiveness of using social media to study public health crises. With regards to the diverse methodologies used to mine social media text, many researchers used NLP-based approaches such as discourse analysis [1], sentiment classification [2], and topic modeling. Social media text mining was utilized to study public health crises related to infectious diseases such as Ebola and Zika. For instance, studies examined Twitter posts from a live chat to identify the public's concerns about the Zika [3] and Ebola [4] epidemics in the United States. The authors in [5] examined the public discourse and emergent themes regarding Malaria by leveraging text mining algorithms, specifically with the use of Crimson social media analytics software.

Concerning the COVID-19 public health crisis, recent studies used twitter data to track or identify public opinions [10], reactions [6], perceptions [7], emotions [8], and concerns [9-13]; to detect topic and sentiment dynamics [14]; to track discourse, mental health status and symptoms [15, 16]; and to detect self-reported symptoms, access to tests, and recovery efforts of users [17]. Twitter has proven to be an effective source for the surveillance of public health data, however, tweets contain limited contextual data due to its character length restriction. As such, other social media platforms have garnered the attention of researchers to exploit context rich data such as Sina Weibo and Reddit. For instance, [18] investigated Chinese citizens' attention to COVID-19 related events by analyzing Sina microblog data. On the other hand, Reddit data have recently been used to observe mental health discourse and COVID-19 related health anxieties [19, 20], to track health related discussions for public health applications [21, 22], to track public COVID-19 concerns [23], and to analyze opioid related discussions by leveraging topic modeling to extract the topics of interest [24].

Topic modeling, specifically Latent Dirichlet Allocation (LDA) and Named-Entity Recognition (NER) are widely used for various text mining tasks for NLP-based social media text analysis [25-28]. Regarding public health, Collier et al. [29] employed topic modeling and NER to detect and track the distribution of infectious disease outbreaks, which demonstrated the feasibility of using both NLP techniques to infer deeper knowledge about a specific health crisis. Similarly, we employed LDA topic modeling to investigate public issues and concerns, and developed a custom NER system to detect the prevalent mitigation strategies by analyzing Reddit data from four North Carolina subreddit communities. To our knowledge, this is the first study to use both LDA topic modeling and a custom NER model on the same Reddit posts to monitor the COVID-19 public health crisis in North Carolina.

## 3 Methodology

We used application programming interfaces (APIs) (Python Reddit API Wrapper[2] and Python Pushshift.io API Wrapper[3]), and a set of predefined search terms ("corona virus", "Coronavirus", "COVID-19" and "SARS-CoV-2") to extract 122,249 comments from 2,319 Reddit posts from four location-specific North Carolina subreddit communities from March 1, 2020 through August 31, 2020. To understand the public uptake of mitigation measures, detection strategies, and public concerns regarding the pandemic in North Carolina, we developed a custom NER model and an LDA-based topic model approach. Our custom NER model identifies key mitigation and detection measures in unstructured Reddit posts using three mitigation named-entity categories: distancing (DIST), disinfectant (DIT), and personal protective equipment (PPE); and two detection named-entity categories: symptoms (SYM) and testing (TEST). Our datasets and code are publically available at https://github.com/NCAT-NLP/ACM-BCB

### 3.1 Dataset Construction

The dataset construction involves data collection from Reddit platform, data preprocessing, building a corpus for topic modeling, and custom NER dataset construction.

*3.1.1 Application Programming Interfaces.* To programmatically access Reddit posts, we used Python Reddit API Wrapper (PRAW), a python package which allows easy access to Reddit's API. After completing the credentialing prerequisites, we gained access to an authorized Reddit instance. In conjunction with PRAW, we used Python Pushshift.io API Wrapper (PSAW), a python package that simplifies passing search parameters that returns public Reddit submissions/comments via the pushshift.io API. In this study, the primary attributes concerning a Reddit/subreddit instance are 'id' and 'name'. Regarding submission objects, the primary attributes are 'id', 'created_utc', 'title', and 'num_comments'. Moreover, 'submission (parent post)', 'body', and 'replies' are the primary attributes for comment objects[2].

*3.1.2 NC Dataset.* To scrape and store the Reddit data, we generated Python scripts. Using our Reddit instance, we selected the criteria for extracting posts and comments containing COVID-19 keywords from March 1st through August 31st 2020 from the subreddits: r/Charlotte, r/raleigh, r/gso (Greensboro), and r/NorthCarolina. The first three subreddits represent communities of the top three populated cities in North Carolina, and r/NorthCarolina represents the subreddit for the entire state. Thus, we created a query selecting posts and comments from the given subreddits with the keywords *corona virus* OR *Coronavirus* OR *COVID-19* OR *SARS-CoV-2*, between the dates 3/1/2020 AND 8/31/2020 (inclusive). Next, we looped through the query results (omitting duplicates) and saved the comment objects to a list. Using the list of comment objects, we then retrieved each comment's parent submission (Reddit post) id, submission date, and submission title. Additionally, we used the submission ids to retrieve the entire comment thread associated with the parent post, including the total number of comments contained in each thread. We then organized the acquired data in a list arranged by date, title, comment body, and number of comments. Upon completion of the list, the data was converted to a table, sorted by date in descending order, and saved to a file in UTF-8 text format. The final composition of our NC_dataset is described in Table 1 below.

---
[2] https://praw.readthedocs.io/en/latest/
[3] https://github.com/dmarx/psaw

**Table 1: Characteristics of NC_dataset**

| Subreddit | #Posts | #Comments | #Sentences | Wordcount |
|---|---|---|---|---|
| Charlotte | 564 | 29,343 | 74,361 | 1,113,915 |
| Raleigh | 784 | 45,402 | 125,706 | 1,863,091 |
| gso (Greensboro) | 140 | 3,821 | 10,270 | 143,464 |
| NorthCarolina | 831 | 43,683 | 119,515 | 1,786,445 |
| *Total* | **2319** | **122,249** | **329,852** | **4,906,915** |

*3.1.3 Data Preprocessing.* Data preprocessing is the process of converting raw data to a useful and efficient format. A description of our preprocessing steps are as follows:

1. **URL removal:** Uniform Resource Locator (URL) does not provide any relevant information and deleting URLs do not significantly affect the text information.
2. **Lowercase:** To ensure all tokens map to the corresponding feature, we convert all the characters to lowercase.
3. **Tokenization:** We employ a sentence tokenizer to break the sentences into individual words.
4. **Stop word removal:** Common, frequently occurring words are removed as they do not add value to the context of the text (e.g., *and*, *the*, *to*, etc.).
5. **Part-of-speech (POS) tagging:** POS tagging is the process of assigning words with their corresponding part of speech tags according to the definition and context (e.g., needs→VERB, needs→NOUN, etc.). It is useful for extracting relations between words and also important lemmatization.
6. **Lemmatization:** Generates the root form of words with knowledge of the context.

Prior to performing LDA topic modeling on the documents from the NC_dataset, the following preprocessing steps were performed: removal of URLs and extra whitespace, tokenization, removal of stop words and digits, POS tagging, lemmatization, removal of non-ASCII text, lowercasing, and the removal of punctuation. The resulting cleaned text was inserted next to the original raw document text. Within the scope of this study, a Document is hereafter defined as the comment thread for each subreddit post. Thus, the number of Documents for each subset from the NC_dataset is equal to the number of posts from each subreddit. Concerning NER, our files that contained our detected named-entities after performing custom NER task were further preprocessed prior to performing a count that combined variants of the same named-entities (e.g., tests, tested, testing → test). Using NLTK, a Python package for statistical natural language processing, we first assigned the named-entities with POS tags.

We then lemmatized the named-entities by which the POS tags determined the word's root form.

*3.1.4 NER Dataset.* Prior to training our custom NER model, we needed a large amount of data similar to the structure of the comment body text of our NC_dataset. Thus, we scraped additional data from the subreddits of three highly populated COVID-19 hotspots as of August 1, 2020. The hotspot subreddits include r/arizona, r/florida, r/texas, r/CoronavirusAZ, r/coronavirusflorida, and r/CoronaVirusTX. To retrieve and store our raw data, we used a modified approach to our NC_dataset construction procedure, differing only in how we stored the data. Once the hotspot subreddit comment objects were stored in a list, we obtained the post comment threads and saved them to a text file in UTF-8 text format. This process was repeated for each of the hotspot subreddits until all the files completed, which were then combined to create our raw text NER corpus, yielding 690,027 sentences.

Using EmEditor[4], data cleaning and editing were performed on the raw NER corpus. First, we used a regular expression to search and remove all URLs. We then searched and removed all instances of the strings '[removed]' and '[deleted]', which are returned as Reddit comment body text if a comment was deleted by a user or removed by a moderator. We removed duplicate sentences before extracting relevant sentences for our NER training and evaluation dataset construction.

*3.1.4.1 Annotated NER Datasets.* Prior to extracting sentences to construct our annotated NER training/evaluation datasets, we first determined five COVID-19 related named-entity categories: personal protective equipment (PPE), disinfectant (DIT), symptoms (SYM), testing (TEST), and distancing (DIST). We then determined keywords within our five entity categories to use for sentence extraction based on the CDC's guidelines for COVID-19 mitigation. A few examples for each include: PPE – face mask, goggles, plexiglass; DIT – soap, bleach, Lysol; SYM – difficulty breathing, diarrhea, fatigue; TEST – false negative, testing site, PCR; and DIST – work from home, social distance, lockdowns.

We first constructed a list of keywords related to our five NER categories to search for the associated sentences. We included up to 250 search results per keyword to our reduced NER corpus. We kept the sentence count to 250 to provide nearly the same amount of training examples for each named-entity. After completing sentence extraction for all five named-entity categories, we removed duplicate sentences. Some sentences contained more than one target keyword which resulted in duplicates when performing our keyword search. Our reduced raw NER corpus now contained 13,829 sentences and we randomly selected 65 percent for our training set (9,040 sentences), and 35 percent for evaluation (4,786 sentences). Before proceeding with the construction of our annotated NER dataset, we replaced the spaces in between all multi-word entities with an underscore (e.g. [social distance] → [social_distance]). The replacement at this phase made it easier to format bigram and n-gram entities prior to the annotation phase. Next, we used a combination of automation and manual configuration to prepare the corpus for labeling.

Using a Python script, we first imported the reduced raw NER corpus. We then tokenized each sentence using spaCy[5], an open-source Python software library for advanced natural language processing, and stored each token in a column labeled 'Words'. Additionally, an adjacent column was created named 'Tags' which we initialized each cell with an 'O' label. This procedure was conducted for both the training and evaluation datasets. The training set yielded 202,309 pre-labeled words and 106,513 words for the evaluation set. The consensus for determining the predefined labels and the manual annotations were conducted by two undergraduate and three graduate-level research assistants.

The named-entity tokens in our annotated datasets were labeled using the BILOU (Beginning, Inside, Last, Outside and Unit) annotation scheme. A BILOU formatted token is assigned B-label if a word is the first token in a multi-word named-entity (label represents one of our predefined entity types, i.e. B-PPE, etc.). A word is assigned I-label if the token is in a named-entity but is not the first or last token. L-label is assigned if a word is the last token in a multi-word named-entity. A word is assigned 'O' if the token is outside a named-entity, and U-label if the token is a single-word named-entity. We initially arranged our underscore-separated multi-word entities on single lines. Entities formatted in the previously described manner are detected as a single entities (bigrams, n-grams) during NER entity detection (e.g. [social DIST distance DIST] → [social distance DIST]). From this point, we labeled each named-entity following the conventional BILOU scheme. It was unnecessary to include the BILU- prefixes next to the entity tags as spaCy would automatically assign said tags with the appropriate BILOU labels during the training/evaluation data format conversion phase. A breakdown of the number of labeled entities in the annotated training/evaluation datasets are shown in Table 2 below.

**Table 2: Number of labeled entity tags**

| Training Dataset | | Evaluation Dataset | |
| --- | --- | --- | --- |
| Entity Label | # of Tags | Entity Label | # of Tags |
| DIST | 1,354 | DIST | 882 |
| TEST | 5,240 | TEST | 1,612 |
| SYM | 3,519 | SYM | 1,574 |
| DIT | 1,562 | DIT | 701 |
| PPE | 3,873 | PPE | 2,010 |
| O | 202,853 | O | 99,236 |

## 3.2 Named-Entity Recognition (NER)

NER is a sub-task of Information Extraction (IE) which aims to seek out and classify certain named-entities found within an unstructured body of a text corpus [30]. Entities found using the NER task are typically classified into pre-defined categories such as person, location, organization, medical codes, disease names, etc. NER is also commonly referred to as entity chunking, entity identification, or entity extraction. NER models are required to be evaluated in order to check the validity of their performance by comparing the outputs typically against human-annotated tags. The comparisons are generally quantified simultaneously by attempting to correctly recognize a detected instance's boundary and its entity type. Consequently, true positives (TP) and true negatives (TN) can be identified to determine when instances are correctly predicted, and false positives (FP) and false negatives (FN) to determine when mistakes occurred. From the aforementioned determinations; precision, recall, and F1 score are assessed. Precision (P) is the percentage of instances the NER classifier got correct out of the total number of instances that it predicted for a given tag. Recall (R) is the percentage of instances the classifier actually predicted for a given tag relative to the number it should have predicted. F1 score (F1) is simply the harmonic mean of precision and recall.

$$P = \frac{TP}{(TP + FP)}, R = \frac{TP}{(TP + FN)}, F1 = 2 \times \frac{P \times R}{P + R}$$

The custom NER models trained for this study were based on spaCy's multi-task convolutional neural network (CNN) which was trained using the OntoNotes[6] corpus, and contain GloVe vectors [31] that were trained on Common Crawl. Prior to inputting the training data to train our NER model, the annotated dataset had to go through several conversions to get the data in the proper spaCy format. Using our annotated training and evaluation NER datasets, we then trained and evaluated three novel, custom NER models. After selecting our highest performing model, we finally performed the custom NER task on the processed Documents from our NC_dataset.

*3.2.1 Training and Evaluation.* Prior to training and evaluating our custom NER models, we created and labeled the training and evaluation datasets as outlined in section 3.1.4.1. To train our custom NER models, we first created predetermined NER labels, provided a model name, and created an NER pipeline. We followed spaCy's recommendations for fine-tuning the following hyperparameters: training iterations, batch size, and dropout rate. A batch size begins at a user defined minimum and each batch increases until it reaches a user defined maximum threshold. Dropout is a stochastic regularization technique that aims to reduce overfitting in neural networks by temporarily removing neurons during training [32]. Custom_NER model 1 was trained with 30 iterations, a compounding batch size from 4 to 32, and a dropout rate of 0.5. Custom_NER model 2 was trained with 50 iterations, a batch size from 1 to 16, and a dropout rate of 0.35. The adjusted hyperparameters used to train Custom_NER model 3 were number of iterations and dropout rate which were 100 and an incremental decay from 0.6 to 0.35, respectively. To evaluate our models, we used spaCy Scorer.

*3.2.2 Entity Detection.* To detect our named-entities, we performed custom NER task on the cleaned dataset from the North Carolina subreddits. We then extracted the named-entities and their corresponding tags and stored them. Due to spaCy's default one-million character limit when parsing text loaded within their NLP document parser, we processed each Document from our NC_dataset one at a time to the document parser, and we appended the detected named-entities and tags to our tables.

---
[4] https://www.emeditor.com/
[5] https://pypi.org/project/spacy/

[6] https://catalog.ldc.upenn.edu/docs/LDC2013T19/OntoNotes-Release-5.0.pdf

### 3.3 Topic Modeling

A topic model is a type of statistical model that efficiently analyzes large volumes of unlabeled text by clustering Documents into topics by discovering hidden semantic structures in a text corpus. Although some researchers have used other topic modeling approaches such as Author-Topic Model (ATM) [33], the most common type of topic model is latent Dirichlet allocation (LDA), which requires the selection of a $k$ value that indicates the number of underlying topics in a text corpus [34]. Initially, LDA assigns each word in a corpus to a random $k$ topic, which is then iteratively updated based on the prevalence of each word across the $k$ topics. Once optimized, LDA employs the Term Frequency-Inverse Document Frequency (tf-idf) metric which assigns probabilities based on the number of occurrences of a word in a Document, and then offsets the probabilities according to the total number of Documents. The topic assignments are continuously updated until a user defined threshold is reached or until the iterations no longer have much impact on the probability assignments to the words in the corpus.

In this study, we designed and implemented our topic models using Python's Scikit-learn library, assigning $k$ = 5 topics to surveil the underlying topics of discussion of North Carolina residents during COVID-19 pandemic from their Reddit posts in four location-specific NC subreddits. Additionally, we obtained the word clouds of the top 15 frequent keywords for each topic and recorded the topic assignment frequency for each Document to aid in topic and theme interpretations.

*3.3.1 Design and Implementation.* Regarding our topic models, features are the words contained in the Documents from our NC_dataset. The feature extraction algorithm that we used for this study was Python's CountVectorizer from the Scikit-learn library. The parameters used were max_df = 0.90 and min_df = 3, by which terms that had a Document frequency strictly higher than 90% were ignored when building the vocabulary, and terms that had a Document frequency strictly lower than 3 were ignored when building the vocabulary, respectively. We then implemented our LDA model using Scikit-learn's LatentDirichletAllocation class. The number of $k$ topics were 5 (n_components) and the number of high frequency keywords were 15. By first conducting three preliminary runs using k = 5, 7, and 10, we determined that k = 5 yielded the most consistent number of coherent topics across all subreddits. Additional parameters used for implementation were learning_method = 'online', learning_offset = 15, and random_state = 42. Once the analysis completed, the size of the unique vocabulary for the combined Documents, and the high probability keywords for each of the $k$ topics were exported to a text file. Additionally, the word clouds for the high probability keywords for each of the $k$ topics were captured and saved as image files. Furthermore, the original datasets were appended with two additional columns that 1) recorded the topic number that each Document belonged to and 2) the probability that each Document was about the assigned LDA topic. The resulting text files, high frequency wordcloud images, and extended datasets were then used to interpret what each of the $k$ = 5 topics represented and their general themes.

## 4 Discussion and Results

This section outlines the results of our study to include general statistics of our NC_dataset, LDA topic modeling and NER analysis, discussions, and limitations of this research.

### 4.1 Distribution of Reddit Posts

Figure 1 shows the frequency of posts by week during our specified date range for each North Carolina subreddit. The week of 3/22/2020 – 3/28/2020 contained the highest number of posts for all subreddits with the exception of r/gso (Greensboro). The largest number of posts for r/gso occurred during the week of 3/15/2020 – 3/21/2020. We observed that the first and second highest number of posts for all subreddits occurred from 3/15/2020 – 3/28/2020.

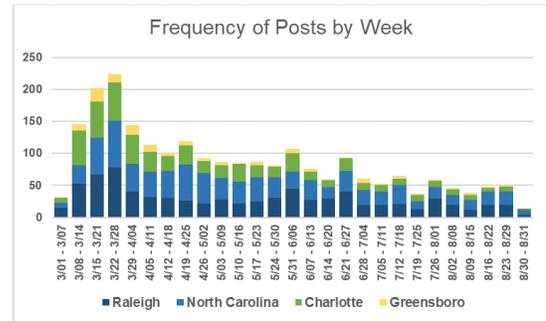

**Figure 1: Number of posts by week from March 1 to August 31, 2020**

**Table 3: NER model performance metrics**

| Model | Precision | Recall | F-score |
|---|---|---|---|
| Custom_NER Model 1 | 92.09 | 90.94 | 91.51 |
| Custom_NER Model 2 | **92.91** | 94.36 | 93.61 |
| Custom_NER Model 3 | 92.74 | **95.20** | **93.95** |

### 4.2 NER Results

*4.2.1 Evaluation Metrics of NER Models.* Using our annotated NER training dataset, we trained three custom NER models by optimizing hyperparameters prior to training. We then evaluated each model using our annotated evaluation dataset. Table 2 depicts the average performance metrics for each model. In terms of F-score, Custom_NER Model 3 achieved the best results. It was further observed that Custom_NER Model 3 outperformed the other models in all metrics except precision, yet negligibly. Accordingly, we selected Custom_NER Model 3 to perform our custom NER analysis. With our chosen model, we conducted NER task on the cleaned, processed Documents from our NC_dataset. Figure 2 is a clustered column chart depicting the number of detected entities for each subreddit. It is shown that the most entities in every category were detected from the Raleigh subreddit. It is additionally shown that the least detected entities in all categories occurred from the Greensboro subreddit. Expectedly, the total number of detected entities for each subreddit mostly correlate with the total number of comments as shown in Table 1 in section 3.1.2.

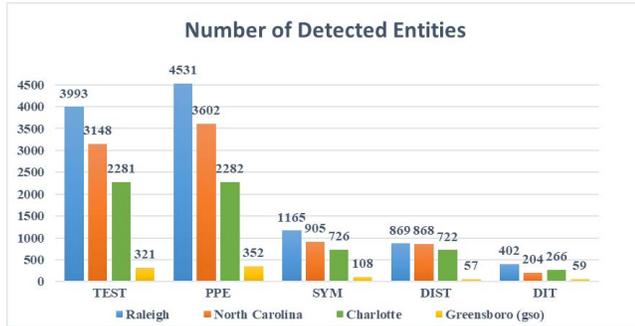

**Figure 2: Number of entities per entity type per subreddit**

*4.2.1 Detected Entity Analysis.* Table 4 lists the top three detected DIST named-entities for each subreddit in addition to the top eight detected named-entities in all other named-entity categories. Across all subreddits, *social distancing, hand sanitizer,* and *mask* were the most prevalent entities for DIST, DIT, and PPE, respectively. Moreover, *flu* and *testing* were the most prevalent COVID-19 detection related entities for the SYM and TEST categories, respectively. S*ocial distancing* dominated as the top distancing mitigation entity for all subreddits. Based on the averages for our representative sample, *social distancing* (57%) and *lockdown* (33.5%) represents 90.5% of the total mitigation entities mentioned across all North Carolina subreddits regarding distancing. In regards to disinfectant (DIT) related COVID-19 mitigation strategies, *hand sanitizer* (25%) and *wipes* (16.25%) indicates 41.25% of the average disinfectant named-entities mentioned across all subreddits. As for PPE related mitigation entities, *mask* denotes an overwhelming average of 85.5% of the prevalent named-entities detected across all subreddits, which is 81.5% more than the second most prevalent PPE entity *gloves* (4% avg.). Concerning symptom (SYM) related COVID-19 detection entities, *flu* makes up nearly half of the most prevalent SYM named-entities detected across all subreddits with an average of 49%. The second most prevalent SYM entity detected across all North Carolina subreddits was *cough*, representing an average of 17.75% SYM entities. For the COVID-19 detection based entity type TEST, *testing* was the named-entity that encompassed the majority of mentions across all North Carolina subreddits with an average of 57.5%. Furthermore, we found that *antibody test* was the most prevalent detected named-entity concerning specific types of COVID-19 detection tests.

This observation is significant as antibody tests can aid with tracking the spread of COVID-19, which according to [35], can lead to more accurate representations of the COVID-19 pandemic.

Although *positive* and *negative* represents the second and third most mentioned TEST entities in all subreddits, it is difficult to readily infer the context in which they were mentioned. For example, clearly there are more COVID-19 negative citizens in North Carolina than there are *positive*, however, *positive* is mentioned about an average of 2.4 times more than *negative*. One could posit that the positive mentions are related to those posting about their or a loved one's COVID-19 status, or some Reddit users could be posting statewide statistics, however, there is no way to know short of manually reading multiple comments.

Subsequently, we determined that it was necessary to conduct further analysis on a particular named-entity to learn the overall discourse based on the manner in which the named-entity is being discussed. Thus, we focused on the most prevalent detected named-entity (*mask*) from the NorthCarolina subreddit given that the results shown in Table 4 are closely aligned with the average results for the subreddits representing the three highest populated cities in the state of North Carolina.

After first extracting all *mask* related sentences (3,917) from the NorthCarolina subreddit, we conducted sentiment analysis using NLTK's sentiment package (i.e. SentimentIntensityAnalyzer) to gauge sentiment. Once duplicate and incomplete sentences were removed, the results from the analysis yielded 1,340 neutral, 1,282 positive, and 1,202 negatively labeled sentences. In solely looking at labels, the comments would lean slightly neutral (positive when disregarding neutral), however, with no indication of strength. Thus, we took the average compound scores of all comments which resulted in a score of 0.007. Since a score of 1 would indicate highly positive and -1 highly negative, it is thus inferred that the comments from the NorthCarolina subreddit based on masks are essentially neutral. Due to the overwhelming neutrality of the comments, we focused on the comments labeled negative to further explore the context in which the users/comments were discussing masks. As we manually inspected the sentences, we begin to notice reoccurring themes and labeled them accordingly as referenced in Table 5.

In terms of social media text and the context in which a certain entity is being discussed, we found that sentiment alone does not necessarily accurately depict the overall discourse or tone. For example, many of the comments that were in fact accurately labeled as negative were about adherence to wearing masks and mask related policies. In these instances, the users' delivery was negative (e.g. "I always wear my mask, these freaking morons should do the same"), yet the discourse was in favor of masks. Moreover, other comments were about businesses and/or law enforcement enforcing the NC mask mandate and/or mask wearing in general. Other contrary themes for comments labeled negative include exemptions and mask efficacy.

Table 4: Identification of entities for 3 mitigation types (distancing, disinfectant, and PPE), and 2 detection types (symptoms and testing). Most distinct and frequently mentioned entities, and their contributing percentages are in bold (Heading represents the four NC subreddits).

| | Charlotte | | | Raleigh | | | Greensboro (gso) | | | NorthCarolina | |
|---|---|---|---|---|---|---|---|---|---|---|---|
| Entity Type | Entity Name | Number of Entities | Entity Type | Entity Name | Number of Entities | Entity Type | Entity Name | Number of Entities | Entity Type | Entity Name | Number of Entities |
| DIST | **social distancing** | 377 (**52%**) | DIST | **social distancing** | 500 (**58%**) | DIST | **social distancing** | 38 (**67%**) | DIST | **social distancing** | 446 (**51%**) |
| | **lockdown** | 262 (**36%**) | | **lockdown** | 293 (**34%**) | | **lockdown** | 13 (**23%**) | | **lockdown** | 360 (**41%**) |
| | work from home | 77 | | work from home | 71 | | work from home | 5 | | work from home | 58 |
| DIT | **hand sanitizer** | 71 (**27%**) | DIT | **hand sanitizer** | 107 (**27%**) | DIT | **hand sanitizer** | 16 (**27%**) | DIT | **hand sanitizer** | 46 (**23%**) |
| | **wipes** | 44 (**17%**) | | **wipes** | 70 (**17%**) | | **wipes** | 11 (**19%**) | | hygiene | 27 |
| | soap | 34 | | soap | 41 | | soap | 8 | | **wipes** | 25 (**12%**) |
| | hygiene | 33 | | bleach | 34 | | hygiene | 6 | | sanitizer | 25 |
| | lysol | 23 | | hygiene | 33 | | clorox | 5 | | bleach | 21 |
| | sanitizer | 23 | | sanitizer | 28 | | bleach | 5 | | soap | 19 |
| | bleach | 12 | | lysol | 20 | | lysol | 2 | | clorox | 11 |
| | clorox | 8 | | clorox | 19 | | alcohol | 2 | | lysol | 10 |
| PPE | **mask** | 1944 (**85%**) | PPE | **mask** | 3900 (**86%**) | PPE | **mask** | 290 (**82%**) | PPE | **mask** | 3188 (**89%**) |
| | **glove** | 115 (**5%**) | | **glove** | 202 (**4%**) | | face mask | 21 | | face mask | 109 |
| | n95 | 80 | | n95 | 115 | | **glove** | 17 (**5%**) | | **glove** | 80 (**2%**) |
| | face mask | 47 | | face mask | 150 | | n95 | 9 | | n95 | 66 |
| | cloth | 42 | | cloth | 82 | | cloth | 5 | | cloth | 67 |
| | respirator | 24 | | respirator | 29 | | face shield | 5 | | respirator | 34 |
| | gown | 11 | | face shield | 13 | | gown | 2 | | face shield | 19 |
| | face shield | 8 | | kn95s | 13 | | respirator | 1 | | gown | 17 |
| SYM | **flu** | 349 (**48%**) | SYM | **flu** | 543 (**47%**) | SYM | **flu** | 49 (**45%**) | SYM | **flu** | 505 (**56%**) |
| | **cough** | 131 (**18%**) | | **cough** | 221 (**19%**) | | **cough** | 18 (**17%**) | | **cough** | 142 (**17%**) |
| | fever | 54 | | fever | 90 | | fever | 15 | | fever | 52 |
| | pain | 44 | | pain | 55 | | pain | 10 | | pain | 50 |
| | cold | 28 | | influenza | 45 | | short of breath | 8 | | influenza | 42 |
| | fatigue | 17 | | short of breath | 33 | | cold | 2 | | headache | 26 |
| | influenza | 17 | | cold | 31 | | difficulty breathe | 2 | | cold | 15 |
| | short of breath | 12 | | headache | 28 | | influenza | 2 | | short of breath | 15 |
| TEST | **testing** | 1382 (**61%**) | TEST | **testing** | 2051 (**51%**) | TEST | **testing** | 190 (**59%**) | TEST | **testing** | 1842 (**59%**) |
| | positive | 498 | | positive | 1122 | | positive | 61 | | positive | 796 |
| | negative | 144 | | negative | 253 | | negative | 22 | | negative | 157 |
| | **antibody test** | 67 (**3%**) | | % positive | 161 | | test kit | 16 | | **antibody test** | 76 (**3%**) |
| | test result | 49 | | **antibody test** | 146 (**4%**) | | **antibody test** | 8 (**2%**) | | test kit | 66 |
| | test kit | 35 | | test result | 54 | | test result | 7 | | % positive | 54 |
| | % positive | 20 | | test kit | 53 | | false positive | 6 | | test result | 46 |
| | false negative | 14 | | false positive | 30 | | test site | 6 | | false positive | 25 |

Conversely, themes that could be deemed neutral or associated with negativity regarding masks include noncompliance, politicizing masks, ineffectiveness, lack of enforcement, sarcastic remarks, selfishness, mask shaming, shortages, skepticism, anti-maskers, wearing masks improperly, hypocrisy, and discomfort.

**Table 5: Reoccurring themes from mask related comments**

| Themes | Frequency |
|---|---|
| noncompliance | 186 |
| compliance | 146 |
| politicized | 117 |
| ineffective | 115 |
| effective | 66 |
| nonenforcement | 63 |
| enforcement | 62 |
| sarcasm | 62 |
| selfishness | 59 |
| mandate | 50 |
| shaming | 43 |
| shortage | 40 |
| skepticism | 37 |
| anti-mask | 32 |
| improper use | 27 |
| hypocrisy | 25 |
| exemptions | 24 |
| uncomfortable | 20 |

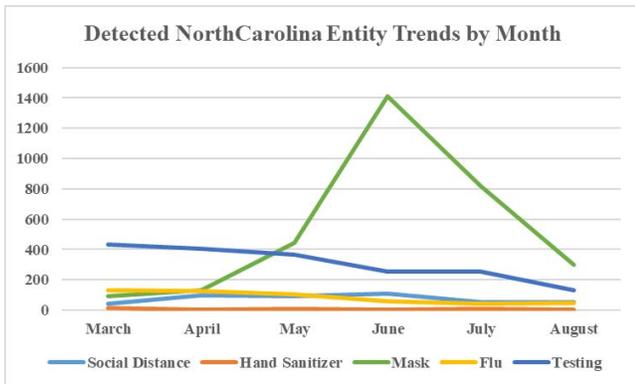

**Figure 3: Month-to-month trends for most prevalent entities.**

The final NER analysis that we conducted was to monitor the evolution of the most prevalent detected entities from the NorthCarolina subreddit dataset from March through August. As shown in Figure 3, *hand sanitizer* is the entity that has the least substantial progression during our observable time span. In fact, distancing (DIT) related entities are the least mentioned entities throughout all datasets. *Flu* and *social distancing* follows similar patterns in terms of month-to-month trends; with social distancing slightly peaking above flu in June. *Testing* is by far the entity that was discussed the most at the onset of the COVID-19 pandemic in the NorthCarolina subreddit and remained fairly steady until May. There was a modest decline from May to June, remained steady until July, then had a comparable decline from July to August as it did from May through June. *Mask* is the most anomalous in terms of prevalently mentioned named-entities from March through August. Mask went from being rarely mentioned in March and April, slightly overtook *testing* in May, and catapulted just over 300% in the following month of June. The decline from June to August was similarly as pronounced as the March to June incline.

### 4.3 Topic Modeling Results

Prior to performing LDA topic modeling, we extracted features from the Documents in our NC_dataset. As a result, the number of unique vocabulary words for each of the subreddits are as follows: r/Charlotte – 10,744, r/raleigh – 14,031, r/gso – 3,143, and r/NorthCarolina – 13,611. Table 6 depicts the interpreted topics and themes for each subreddit and Figure 4 shows the word cloud representing the 15 high frequency words used to shape each of the $k = 5$ topics.

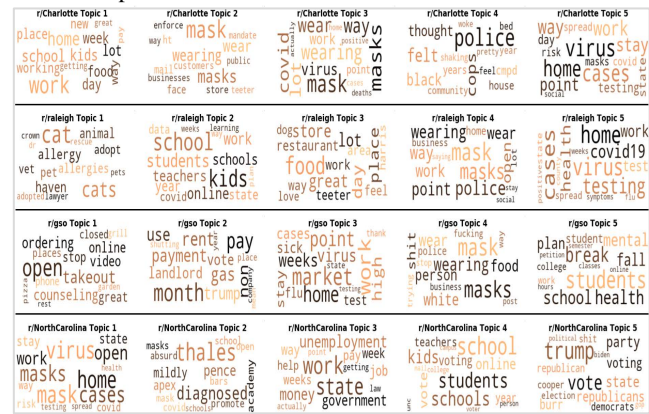

**Figure 4: Word clouds of the top 15 frequent keywords for each topic (use Table 6 for indexing).**

We found that in nearly all cases, the top 15 frequent keywords sufficiently shaped clearly defined topics. In few instances however, topics seemed to overlap and/or needed a deeper analysis for topic interpretation. Concerning r/Charlotte, it is clearly shown in Topic 1, *work* and *home* are two of the most frequently occurring keywords, hence *Work from Home* being selected as a likely candidate as primary topic. Words such as *working*, *week*, *day*, and *pay* seem to shape a secondary topic. As mentioned in section 3.3, we assigned all Documents with the topic number that LDA determined each Document was about, coupled with the topic's corresponding probability. To determine secondary themes or to verify loosely defined topics, we randomly selected Documents assigned a specific topic that when possible, had probability scores 90% or greater. An analysis of the randomly selected Documents (RSD) revealed that the keywords related to Topic 1's potential secondary theme was *jobs*. Another interesting observation regarding r/Charlotte was related to its second and third topic. In both Topic 2 and Topic 3, *wearing*, *mask*, and *masks* were three of the most frequently occurring words, which indicated the primary topic as *Wearing Masks*. It seemingly appeared as if the LDA topic modeling captured the same topic twice, which may be true in some regard.

**Table 6: Interpreted LDA topics with categories**

| Subedit Name | Topic Number | Topic Name | Category |
|---|---|---|---|
| r/Charlotte | 1 | Jobs / Work From Home | Employment |
|  | 2 | Wearing Masks (mandate) | COVID-19 Mitigation Measures |
|  | 3 | Wearing Masks | COVID-19 Mitigation Measures |
|  | 4 | Social Justice/Law Enforcement | Law Enforcement |
|  | 5 | COVID-19 testing/cases/measures | Spread of COVID-19 Disease |
| r/raleigh | 1 | Pets | Pets |
|  | 2 | Education | Education |
|  | 3 | Shopping | Business |
|  | 4 | Wearing Masks | COVID-19 Mitigation Measures |
|  | 5 | COVID-19 testing/cases/measures | Spread of COVID-19 Disease |
| r/gso (Greensboro) | 1 | Dining | Business |
|  | 2 | Housing | Housing |
|  | 3 | COVID-19 testing/cases/measures | Spread of COVID-19 Disease |
|  | 4 | Wearing Masks | COVID-19 Mitigation Measures |
|  | 5 | Education | Education |
| r/NorthCarolina | 1 | Masks / COVID-19 testing/cases/measures | COVID-19 Mitigation Measures / Spread of COVID-19 Disease |
|  | 2 | Thales Academy | Education |
|  | 3 | Unemployment | Employment |
|  | 4 | Education | Education |
|  | 5 | Politics/2020 Election | Politics |

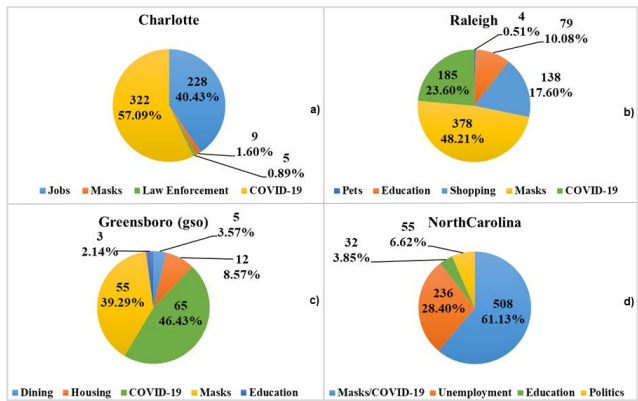

**Figure 5(a)-(d): Number of assigned topics and their overall discussion percentages. Refer to Figure 4 and Table 6 for indexing.**

However, we further observed that zero occurrences were assigned by LDA for Topic 3 in r/Charlotte. Consequently, Topic 3 was disregarded and Topic 2 was further evaluated. A closer analysis of r/Charlotte's Topic 2 revealed the context in which wearing masks were discussed. It was found that the discussions primarily dealt with mask enforcement or the North Carolina mask mandate. In terms of topic frequency, *COVID-19 testing/cases/measures* (57%) was the most discussed topic in r/Charlotte followed by *Jobs / Work from Home* (40%) and *Wearing Masks* (2%). *Social justice* (1%) was the least discussed. Combined, *COVID-19 testing/cases/measures* and *Jobs / Work from Home* represented 97% of the most discussed topics in r/Charlotte. It is also worth noting that the keyword *masks* encompassed *COVID-19 testing/cases/measures*, thus inferring the context of masks being discussed primarily as *COVID-19 Mitigation Measures*.

Regarding r/raleigh, Topic 3 was the only topic that was not readily apparent when considering only the 15 most frequently occurring keywords. As shown in Figure 5(b), Topic 3 was the third most discussed topic, significantly contributing to the overall discussion for r/raleigh. Therefore, we conducted a deeper analysis of Topic 3 using the aforementioned RSD method. The keywords *harris*, *teeter*, *food*, *restaurant*, and *store* were good indicators of a coherent topic, although loose. The deeper analysis revealed it to be *Shopping*. Although insignificant in relation to topic frequency, we were surprised that *Pets* (Topic 1) was rendered as a well-defined topic of discussion during the COVID-19 pandemic. RSD revealed that the pet related discussions were in the context of adopting or treating pets 'during' the COVID-19 pandemic rather than COVID-19 'directly' affecting pets. As a whole, we found that *Wearing Masks* (48%) was the most discussed topic in r/raleigh followed by *COVID-19 testing/cases/measures* (24%), *Shopping* (18%), and *Education* (10%). *Pets* was the least discussed topic, representing less than 1% of the overall discussions.

The subreddit r/gso had the least amount of posts and comments, however, LDA topic modeling was very effective in shaping well defined, coherent topics which made interpreting topics for r/gso relatively straightforward. An interesting observation was the emergence of two topics independent of the other subreddits which were *Dining* (Topic 1), and *Housing* (Topic 2). Figure 5(c) shows that housing was the third most discussed topic and dining was fourth. The results were significant enough to infer that the residents of Greensboro were concerned about how the pandemic affected their dining experiences and their housing situations. The most discussed topic in r/gso was *COVID-19 testing/cases/measures* (46%) followed by *Wearing*

*Masks* (39%), *housing* (9%), and *dining* (4%). The least discussed topic in r/gso was *Education* (2%).

Regarding r/NorthCarolina, *Thales Academy* (Topic 2) was the topic that needed a deeper analysis. By applying RSD to Topic 2, we discovered that a COVID-19 related incident occurred at one of eight Thales Academy North Carolina campuses, in which several students and a visitor tested positive for COVID-19 at the Knightdale location. Moreover, Vice President Pence visited the Apex campus to discuss the importance of reopening schools amidst the COVID-19 pandemic. This is another example of how the application of RSD can further aid in the discovery of the context in which a clearly defined LDA topic is discussed. As evident by its absence in Figure 5(d), *Thales Academy* (Topic 2) represented zero of the prevalent topics discussed in r/NorthCarolina. Thus, it was observed that RSD can additionally reveal the predominant topic of a Document that contains keywords from unassigned LDA topics. For example, three Documents in r/NorthCarolina were about Thales Academy based on a keyword search, however, all three were assigned *COVID-19 testing/cases/measures* (Topic 1) with an average probability of 69%. We attribute this to the fact that the Thales Academy incidents were in large about COVID-19, thus taking precedence. Concerning *Masks / COVID-19 testing/cases/measures* (Topic 1), the keywords *mask* and *masks* were as prevalent as the *Spread of COVID-19 Disease* related keywords, thus, we inferred that Topic 1 was nearly equally about masks as it was about the spread of the COVID-19 disease. In all other subreddits, *Wearing Masks* was a distinguishable topic separate from *Spread of COVID-19 Disease*-related topics. Regarding the number of assigned prevalent topics discussed in r/NorthCarolina, *Masks / COVID-19 testing/cases/measures* (61%) was the highest followed by *Unemployment* (28%) and *Politics* (7%). The lowest measurable topic discussed in r/NorthCarolina was *Education* (4%).

To finalize our LDA topic modeling analysis, we used our NorthCarolina subreddit dataset as representation for the state of North Carolina to determine the most prevalent side topics discussed in North Carolina on a per month basis. Our impression was that by dividing the dataset into months and performing LDA topic modeling with $k = 2$ topics on each, then one topic would encompass COVID-19 related keywords and the other would produce the most prevalent side topic. This proved true in all but two cases where instead of getting a COVID-19 related topic for the month of May, we got a topic that was mostly incoherent. The other instance was July where two COVID-19 related topics emerged, however, one of them had a distinguishable co-occurring topic (Education). Disregarding the incoherent and COVID-19 related topics, the results for the prevalent side topics on a month-to-month basis are shown in Table 7 and Figure 6.

It was interesting to observe how the predominant side topics of discussion changed from month-to-month, and how they closely corresponded with real-time events that occurred in North Carolina during those months. For example, in March of 2020, Senator Burr of NC was the center of the 2020 congressional insider trading scandal probe, April was the month North Carolinians could begin applying for COVID-19 pandemic unemployment benefits, June was the month that the George Floyd protests began in North Carolina, July is when Governor Cooper announced school reopenings, and August was the month when President Trump encouraged North Carolina supporters to vote with absentee ballots with his mail-in voting messaging due to the surge of COVID-19.

**Table 7: Dominant NC subreddit monthly side topics**

| Month | Topic Name |
| --- | --- |
| March | Coronavirus Insider Trading Scandal |
| April | Unemployment |
| May | Unemployment |
| June | Law Enforcement/Social Justice |
| July | Education |
| August | 2020 Election/Early Voting |

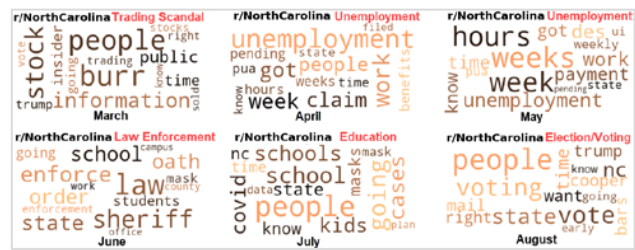

**Figure 6: Top 15 frequency words associated with Table 7**

### 4.4 Discussions

Our multi-pronged NLP study proved highly effective in surveilling the public uptake of mitigation/detection strategies and public concerns regarding the COVID-19 pandemic. As North Carolina based researchers, a good first step was to apply our methodologies on a dataset derived from location-specific North Carolina Reddit data. An advantage of focusing on three location-specific subreddits of a particular state, specifically the three highest populated cities, and by then focusing on the state-specific subreddit, it can be determined whether those cities are a good representative sample for the entire state. Additionally, by focusing on multiple cities, various mitigation/detection uptake trends and concerns specific to those communities can be determined. Furthermore, our comprehensive dataset construction method produces a cleaned corpus which our proposed NLP experiments can be applied on any region that has location-specific Reddit communities to surveil the COVID-19 pandemic.

In the case of custom NER, the efforts of determining categories, extracting relevant sentences, annotation, training and evaluating models are advantageous. It creates a novel, custom NER model unique to researchers' needs and can readily be applied to any appropriately formatted dataset. In this study, our custom NER model effectively detected predetermined COVID-19 mitigation and detection-based entities from four location-specific North Carolina based subreddits, and helped us quickly gauge the most prevalent named-entities for each category across our selected subreddits. Another advantage of using a custom trained NER model is its ability to detect misspelled variations of targeted entities.

As evidenced by this study, the most prevalent entities can be identified using custom NER, which can aid in identifying sentences from the constructed dataset to extract and perform further analysis (e.g. determining the sentiment and the discourse in which the named-entities are being discussed). We noted that *mask* was the most prevalent named-entity discussed in North Carolina and had the most substantial month-to-month progression. It was further noted that the mask related comments were essentially neutral and the discourse of the negatively marked comments were categorized by 18 reoccurring themes, in which the second most reoccurring theme pertained to the adherence of wearing masks.

The LDA topic modeling portion of this study also proved effective in determining the primary concerns of North Carolinians during the COVID-19 public health crisis and there are several advantages with our LDA topic modeling method. First, appending the cleaned dataset with topic numbers and discussion percentages assist with efficiently performing further analysis. For instance, comments associated with a particular topic can be promptly identified to clarify topic names that are not readily interpretable, and insight can be conveniently be gained on the context in which a particular topic is being discussed. Also, topic frequencies and discussion strength percentages can be easily calculated. Other advantages of our LDA topic model methods are that datasets created from multiple subreddits can be analyzed successively, can be processed without interruption due to insufficient number of posts, and are automatically arranged to a directory unique to each subreddit including their respective text files, appended cleaned datasets, and top frequency keyword images.

As shown in this study, our LDA topic modeling method successfully captured concerns from North Carolinians regarding the COVID-19 public health crisis. It was shown that the top discussion percentages in terms of topic frequency was related to COVID-19 mitigation and detection strategies. In similar fashion with our custom NER analysis, our LDA analysis also showed that mask was the most prevalent, specified COVID-19 mitigation strategy, both as a singular topic and through co-occurring keywords that were interpreted as *COVID-19 testing/cases/measures* topics. Furthermore, our LDA method detected interesting side topics related to unexpected consequences of the COVID-19 pandemic. For example, the citizens of Charlotte faced a heavier burden on jobs than in Raleigh or Greensboro, and the pandemic significantly impacted shopping and institutions of higher learning in Raleigh. The most significant indirect impacts of the COVID-19 pandemic in the subreddits of Greensboro and North Carolina were related to housing and unemployment, respectively. It was additionally shown that our LDA method was effective in revealing the most prevalent side topics in the North Carolina subreddit on a month-to-month basis, heavily corresponding with real-time events that occurred in North Carolina during those months.

## 4.5 Limitations

This study is not without a few limitations. First, our sentiment analysis was conducted using an out-of-the-box system. Our manual annotator randomly labeled a small portion of our mask related sentences with pos, neg, and neu which led to performance metric scores of 100% for precision, recall, and F1-score. Thus, the true performance of the sentiment analysis system is unknown. Additionally, only the mask related sentences labeled 'neg' by the system were categorized with reoccurring themes. The first limitation will be addressed in future studies by recruiting several research assistants to manually label all sentences with the appropriate sentiment to more accurately determine system performance. The latter will be addressed by categorizing all prevalent entity-related sentences with reoccurring themes, regardless of sentiment, to better determine the overall discourse. It should however be noted that the primary NLP techniques for this study was LDA topic modeling and custom NER, and the additional sentiment and discourse techniques were used as proof of concepts to strengthen our NER results as NER does not readily infer any sentimental or contextual information.

## 5 Conclusion

COVID-19 pandemic is the latest health crisis currently afflicting the world. As a result, the lives of the global public has drastically been altered. The uncertainties surrounding the COVID-19 global pandemic has led to life altering policies and guidelines that have triggered myriad concerns from global citizens. Subsequently, many people have turned to social media to express their concerns regarding the pandemic, which in turn has provided a massive data repository that researchers can analyze to discover meaningful data, in the hope of combating the spread of the COVID-19 disease.

Today, people around the world are still suffering from the effects of COVID-19. By analyzing social media (Reddit) posts to monitor public uptake of mitigation strategies and concerns about COVID-19 pandemic, the findings of this research can help governments develop the next steps in response, advise the public on protection in personal life and at work, and provide experience for future epidemic prevention based on the mining of social media platforms. By using NLP on Reddit posts, we monitored and demonstrated public concerns about mitigation strategies about the COVID-19 pandemic. Our findings demonstrate the concerns of people in four largest subreddit communities in North Carolina about COVID-19 and related topics.

In this study, we used six months of Reddit data to survey the COVID-19 pandemic in North Carolina by employing LDA topic modeling and a custom NER system with much success. Based on our sample population, we found that the North Carolina public is most concerned with mitigating the spread of COVID-19 by wearing masks, adhering to social distancing guidelines, and using hand sanitizer. We further observed that the most prevalent topics of discussion among the North Carolina public were wearing masks and the spread of COVID-19. Our experiments have demonstrated 1) the effectiveness of using a custom NER system on Reddit posts to monitor prevalent COVID-19 detection and the uptake of mitigation strategies in North Carolina, and 2) the efficacy of employing LDA topic modeling to discover the underlying concerns of North Carolinians during COVID-19 pandemic using Reddit posts.